\documentclass[preprint,showpacs,preprintnumbers,amsmath,amssymb]{revtex4}
\usepackage{graphicx}
\usepackage{bm}
\usepackage{epsfig}
\begin{document}
\title{\bf{Scaling ansatz, four zero Yukawa textures and large $\theta_{13}$}}
\author{ Biswajit Adhikary} 
\affiliation{Department of Physics,Gurudas College,Narkeldanga,
Kolkata-700054, India}
\author{Mainak Chakraborty}
\affiliation{Saha Institute of Nuclear Physics, 1/AF Bidhannagar,
  Kolkata 700064, India} 
\author{Ambar Ghosal}
\affiliation{Saha Institute of Nuclear Physics, 1/AF Bidhannagar,
  Kolkata 700064, India} 
\begin{abstract}
We investigate 'Scaling ansatz' in the neutrino sector 
within the 
framework of type I seesaw mechanism with diagonal charged lepton and 
right handed Majorana neutrino mass matrices ($M_R$). We also assume four 
zero texture of Dirac neutrino mass matrices ($m_D$) 
which severely constrain
the phenomenological outcomes of such scheme. 
Scaling ansatz and the 
present neutrino data allow only Six such matrices out 
of 126 four zero Yukawa matrices. In this scheme, 
in order to generate large 
$\theta_{13}$ we break scaling ansatz in $m_D$ through a perturbation 
parameter and we also show our breaking scheme is radiatively 
stable. We further investigate 
CP violation and baryogenesis via leptogenesis in those surviving textures. 
\end{abstract}
\maketitle
\newpage
\section{Introduction}
\noindent
Neutrino physics is now playing a pivotal role to probe physics beyond the 
Standard Model. Confirmation of tiny neutrino masses as well as nonzero
mixing angles have thrown light on the structure of the leptonic sector. 
In the quest towards understanding of a viable texture of neutrino mass matrix 
popular paradigm is to advocate flavor symmetries, 
directly associated with some gauge group.
On the other hand, there are some other ansatzs which also give 
rise to interesting phenomenological consequences, although their
origin from a symmetry discrete or continuous are yet to be 
established at the present moment.
\vspace{3mm}\\
In the present work we bring together two ideas to explore the 
neutrino phenomenology, particularly, to generate large $\theta_{13}$ 
\cite{Gluza}-\cite{ema}
as reported by recent experiments \cite{minos1}-\cite{last} as well 
as CP violation and baryogenesis via leptogenesis.
In this scheme we consider 
\vskip 0.1in
\noindent
a) Scaling ansatz\cite{sc1}-\cite{sc3}, 
\vskip 0.1in
\noindent
b) Four zero texture 
\cite{4zero1}-\cite{4zero4}
of Dirac neutrino matrix $(m_D)$, 
\vskip 0.1in
\noindent
within the framework of type I seesaw mechanism denoted as 
\begin{equation}
m_\nu = - m_D M_R^{-1}m_D^T
\end{equation}
where $M_R$ is a $3\times3$ right 
chiral neutrino mass matrix and we consider  
the basis in which charged lepton and $M_R$ are flavor diagonal.  
\vskip 0.1in
\noindent
Scaling ansatz \cite{sc1}-\cite{sc3} posses a distinctive feature that the texture 
is invariant under renormalisation group evolution unlike 
other symmetries such as $\mu -\tau$ symmetry. 
Basically, the ansatz correlates the elements of neutrino mass matrix through 
a scale factor and it can be implemented in different ways.
Although the theoretical origin of such ansatz is not yet well known, 
however, this ansatz can be approximated as $S_{2L}$ symmetry (i.e $\mu-\tau$ symmetry in the left handed 
neutrinos) with the 
value of the scale factor unity. 
Furthermore, it leads to inverted hierarchy of neutrino mass 
with $m_3$ = 0, and $\theta_{13}$ = 0. 
Thus, it is obvious to break such 
ansatz  in order to generate nonzero $\theta_{13}$. 
\vskip 0.1in
\noindent
The other assumption that occurrence of four zeroes 
in $m_D$ gives rise to  a more constrained feature that the phases 
contributing to the high scale CP violation 
required for leptogenesis 
(basically the phases of $m_{D}m_{D}^{\dagger}$ matrix) are determined in terms of the low energy CP violating phases
 ( i.e phases of $m_{\nu}$).
\vskip 0.1in
\noindent
We divide all the $9_{{C}_{4}}=126$ four zero textures in the 
following Classes :
\vskip 0.1in
\noindent
i) det($m_D$)= 0  and no generation decouples: $27$ textures
\vskip 0.1in
\noindent
ii) det($m_D$) $\neq$ 0 and no generation decouples: $72$ textures.
\vskip 0.1in
\noindent
iii)det($m_D$)= 0  and one generation decouples: $18$ textures
\vskip 0.1in
\noindent
iv) det($m_D$) $\neq 0$ and one generation decouples: $9$ textures
\vskip 0.1in
The textures belong to Class (ii) are already studied extensively
\cite{4zero1}-\cite{4zero4}. Class (iii) and (iv) are incompatible with
the neutrino experimental result.
The remaining Class,  
Class (i), which is yet to be explored, posses one zero eigenvalue which is still allowed by the 
present experiments.
The interesting point is to note that if we insert scaling ansatz 
to all four zero textures and
consider those textures in which four zero 
remain four zero and no generation decouples, we  see
 that the survived textures  are only from Class (i). 
Motivated with this unique selection property of scaling ansatz,
in the present work we investigate textures belong
to Class (i). 
In addition to one eigenvalue zero, 
scaling ansatz also dictates one mixing angle to be zero. We further
generate nonzero $\theta_{13}$ through the breaking
of scaling ansatz due to a small perturbation parameter in $m_D$. We 
investigate
all possible cases and finally we demonstrate that the broken
scaling ansatz textures remain invariant under renormalization group (RG) 
evolution.
\vspace{3mm}\\
Our plan of this paper is as follows : In Section II we discuss 
different types of scaling ansatz and allowed four zero textures. Section III
 contains parametrization and 
diagonalisation of neutrino mass matrix. 
Breaking of scaling ansatz and generation of nonzero
 $\theta_{13}$ are discussed in Section IV. Numerical results are given in 
Section V and Section VI contains the possible baryogenesis via 
leptogenesis scenario arises in those textures and 
summary of the present work is given in Section VII.  
Discussion on RG effect is given in Apendix A and 
explicit expressions  
arise in Section IV are included in Appendix B. 
\section{Four zero Yukawa textures and Scaling ansatz}
\subsection{Scaling ansatz}
Several authors \cite{sc1}-\cite{sc3}  have been studied scaling ansatz through 
its implementation along the columns of effective $m_{\nu}$ matrix.
 In the present work, we consider this ansatz at a more 
fundamental level of $m_{D}$ \cite{sc1} and we find that implementation of this
ansatz along the rows of $m_{D}$ with a diagonal $M_{R}$ 
effectively gives rise to the same structure of $m_{\nu}$ \cite{sc2} 
 after invoking
type-I seesaw mechanism.
According to this ansatz elements of a row (of 3$\times$3 $m_{D}$) are 
connected with the elements of another row through a
definite scale factor.  
In case of 3$\times$3 $m_{D}$ there are three types of 
this ansatz which are given as follows:\\
 i$\rangle$ Second and third row are related through a complex
scale factor $k$ as
\begin{equation}
{m_{D}}_{\mu i}=k{m_{D}}_{\tau i} \label{s1}
\end{equation}
where $i$ is column index, $i= 1,~2,~3$.
Invoking type I seesaw mechanism
\begin{eqnarray}
{(m_\nu)}_{\mu\alpha}& = &-{(m_D)}_{\mu j}M_{Rj}^{-1}{m_D^T}_{j\alpha}\nonumber\\
      & = & -k{(m_D)}_{\tau j}M_{Rj}^{-1}{m_D^T}_{j\alpha}\nonumber\\  
&  = &  k{(m_\nu)}_{\tau\alpha} \label{ex1}
\end{eqnarray}
with $\alpha=e,~\mu,~\tau$ we obtain the following scaling relations in $m_\nu$
\begin{equation}
\frac{ (m_\nu)_{\mu e}}{(m_\nu)_{\tau e}}=\frac{(m_\nu)_{\mu\mu}}{(m_\nu)_{\tau\mu}}=
\frac{(m_\nu)_{\mu\tau}}{(m_\nu)_{\tau\tau}} = k 
\label{a1}\end{equation}
\noindent
We discard the other two cases where the scale factor relates 
ii$\rangle$ First and third row 
and iii$\rangle$ First and second row  because in those cases 
either $\theta_{12}$ 
or $\theta_{23}$ is zero at the leading order.  
\subsection{Four zero Yukawa textures}
We start with a general scaling ansatz invariant $m_D$ 
matrix on which we will 
assume four zeroes and explore all the possibilities. 
Explicit structure of $m_D$ according to eqn.(\ref{s1}) is given by  
\begin{eqnarray}
m_D = \begin{pmatrix}   a_1 & a_2 &a_3\cr
                        kb_1& kb_2 & kb_3\cr
                        b_1 & b_2 &b_3
\end{pmatrix}
\label{mD}
\end{eqnarray}
\begin{table}[htb]
\begin{tabular}{|c|c|c|}
\hline
\multicolumn{3}{|c|}{{\bf Category $A$}}\\
\hline
$b_1$ = 0 and $a_1$ = $a_2$ = 0 & $b_1$ = 0 and $a_1$ = $a_3$ = 0 & $b_2$ = 0 and $a_1$ = $a_2$ = 0  \\
 $m_D^{IA}=\begin{pmatrix}0 & 0 &a_3\\ 
             0 & kb_2 & kb_3\\
             0 & b_2  & b_3\end{pmatrix}$  
& $m_D^{IIA}=\begin{pmatrix} 0 & a_2 &0\\
    0& kb_2 & kb_3\\
    0 & b_2 &b_3\end{pmatrix}
  $ 
& $m_D^{IIIA} =\begin{pmatrix}0 & 0 &a_{3}\\
                  kb_{1} & 0 & kb_3\\
                  b_{1} & 0 & b_3\end{pmatrix}$\\
\hline 
$b_2$ = 0 and $a_2$ = $a_3$ = 0 & $b_3$ = 0 and $a_1$ = $a_3$ = 0 & $b_3$ = 0 and $a_2$ = $a_3$ = 0  \\
$m_D^{IVA}=\begin{pmatrix}a_{1} & 0 &0\\ 
             kb_{1} & 0 & kb_3\\
             b_{1} & 0  & b_3\end{pmatrix}$ 
& $m_D^{VA}=\begin{pmatrix}0 & a_{2} &0\\
                  kb_{1}& kb_{2} & 0\\
                   b_{1} & b_{2} & 0
                  \end{pmatrix}$
&  $m_D^{VIA}=\begin{pmatrix}a_{1} & 0 &0\\
                    kb_{1}& kb_{2} & 0\\
                     b_{1} & b_{2} &0
                    \end{pmatrix}$ \\
\hline
\multicolumn{3}{|c|}{{\bf Category $B$}}\\
\hline
$b_{1}=b_{2}=0$ & $b_{1}=b_{3}=0$ & $b_{2}=b_{3}=0$\\
$m_D^{IB}=\begin{pmatrix}a_{1} & a_{2} & a_{3}\\ 
                0 & 0 & kb_3\\
                0 & 0  & b_3\end{pmatrix}$
& $m_D^{IIB}=\begin{pmatrix} a_{1} & a_{2} & a_{3}\\ 
                0 & kb_{2} & 0\\
                0 & b_{2}  & 0\end{pmatrix}$
& $m_D^{IIIB}=\begin{pmatrix}a_{1} & a_{2} & a_{3}\\ 
                kb_{1} & 0 & 0\\
                b_{1} & 0  & 0\end{pmatrix}$ \\
\hline
\multicolumn{3}{|c|}{{\bf Category $C$}}\\
\hline
$a_{2}=a_{3}=0$, $b_{1}=0$ & $a_{1}=a_{3}=0$, $b_{2}=0$ & $a_{1}=a_{2}=0$, $b_{3}=0$\\
 $m_D^{IC}=\begin{pmatrix}a_{1} & 0 &0\\ 
             0 & kb_{2} & kb_3\\
             0 & b_{2}  & b_3\end{pmatrix}$
& $m_D^{IIC}=\begin{pmatrix}0 & a_{2} &0\\ 
             kb_{1} & 0 & kb_3\\
             b_{1} & 0  & b_3\end{pmatrix}$
&  $m_D^{IIIC}=\begin{pmatrix}0 & 0 & a_{1}\\ 
             kb_{1} & kb_{2} & 0\\
             b_{1} & b_{2}  & 0\end{pmatrix}$ \\
\hline 
\end{tabular}
\caption{ Four zero Yukawa textures compatible with Scaling ansatz} 
\end{table}
We categorise all possible four zero textures compatible with Scaling ansatz 
in three different cases as shown in Table I. 
The following points to be noted :
\vskip 0.1in
\noindent
\begin{enumerate}
\item {We find that out of 126 four zero textures, 
imposition of  scaling ansatz 
reduces drastically the 
number to only 12.}
\vskip 0.1in
\noindent
\item {We ignore Category B because it is not possible to break 
scaling ansatz keeping the pattern of $m_D$ matrices unaltered. 
Let us assume the breaking is incorporated as 
$k\rightarrow k(1+\epsilon)$, the structure of all $m_D$ remain 
same and still invariant under scaling ansatz. Thus, to break 
scaling ansatz in Category B, we have to have reduce the number 
of zeroes which is beyond our proposition.}
\vskip 0.1in
\noindent
\item {We also discard all the textures in Category C  
since one generation is completely
decoupled
 from the other two which give rise to two mixing angles zero.}
\end{enumerate} 
\noindent
Hence, the number of surviving texture is only six and all of them are 
from Class (i) described previously in the Section I. 
\vskip 0.1in
\noindent
For Category A as the second and third row of the matrices are
connected through a scale factor, from now on we express them as follows 
\vskip 0.1in
\noindent
\begin{eqnarray}
m_D^{IA} = 
\left( \begin{array}{ccc} 0 & a & 0 \\ 0 & kb & kc \\ 0 & b & c \end{array}
\right),
m_D^{IIA} = 
\left( \begin{array}{ccc} 0 & 0 & a \\ 0 & kb & kc \\ 0 & b & c \end{array}
\right),
m_D^{IIIA} = 
\left( \begin{array}{ccc} a & 0 & 0 \\ kb & 0 & kc \\ b & 0 & c \end{array}
\right),
\end{eqnarray}
\begin{eqnarray}
m_D^{IVA} = 
\left( \begin{array}{ccc} 0 & 0 & a \\ kb & 0 & kc \\ b & 0 & c  \end{array}
\right),
m_D^{VA} = 
\left( \begin{array}{ccc} a & 0 & 0 \\ kb & kc & 0 \\ b & c & 0 \end{array}
\right),
m_D^{VIA} = 
\left( \begin{array}{ccc} 0 & a & 0 \\ kb & kc & 0 \\ b & c & 0  
\end{array}\right)\label{col0}
\end{eqnarray}
where $a$, $b$, $c$ and $k$ are all complex parameters.
\section{ Parametrization and Diagonalisation}
\subsection{Parametrization}
\vskip 0.1in
\noindent
We parametrize the 
$m_\nu$ matrix  arises after seesaw for all 
 $m_D$ matrices in Category A in a generic way as 
\begin{equation}
m_\nu=m_0\left( \begin{array}{ccc} 1 & kpe^{i\theta} & pe^{i\theta} \\ kpe^{i\theta} & 
k^2(q^2e^{2i\beta}+p^2e^{2i\theta}) & k(q^2e^{2i\beta}+p^2e^{2i\theta}) \\ pe^{i\theta} & k(q^2e^{2i\beta}+p^2e^{2i\theta}) & 
q^2e^{2i\beta}+p^2e^{2i\theta} \end{array}\right)
\label{mnu1}
\end{equation}
with the definitions of the parameters for six consecutive cases as 
\begin{eqnarray}
m_D^{IA}:\quad\quad 
m_0=-\frac{a^2}{M_2},\quad pe^{i\theta}=\frac{b}{a},\quad qe^{i\beta}=
\sqrt{\frac{M_2}{M_3}}\frac{c}{a}\nonumber\\
m_D^{IIA}:\quad\quad 
m_0=-\frac{a^2}{M_3},\quad pe^{i\theta}=\frac{c}{a},\quad qe^{i\beta}=\sqrt{\frac{M_3}{M_2}} \frac{b}{a}\nonumber\\
m_D^{IIIA}:\quad\quad 
m_0=-\frac{a^2}{M_1},\quad pe^{i\theta}=\frac{b}{a},\quad qe^{i\beta}=\sqrt{\frac{M_1}{M_3}} \frac{c}{a}\nonumber\\
m_D^{IVA}:\quad\quad 
m_0=-\frac{a^2}{M_3},\quad pe^{i\theta}=\frac{c}{a},\quad qe^{i\beta}=\sqrt{\frac{M_3}{M_1}} \frac{b}{a}\nonumber\\
m_D^{VA}:\quad\quad 
m_0=-\frac{a^2}{M_1},\quad pe^{i\theta}=\frac{b}{a},\quad qe^{i\beta}=\sqrt{\frac{M_1}{M_2}} \frac{c}{a}\nonumber\\
m_D^{VIA}:\quad\quad 
m_0=-\frac{a^2}{M_2},\quad pe^{i\theta}=\frac{c}{a},
\quad qe^{i\beta}=\sqrt{\frac{M_2}{M_1}} \frac{b}{a}.
\label{def}
\end{eqnarray}
Considering complex $k$ as $ke^{i\theta}$ and $m_0$ as $m_0e^{i\theta_m}$,
we rotate the matrix $m_\nu$ by 
$e^{-i\theta_m/2}\times{\rm diag}(1,e^{-i(\theta+\theta_k)}, e^{-i\theta})$ 
from both sides 
and get the $m_\nu$ free from redundant phases as 
\begin{eqnarray}
m_\nu=m_0\left( \begin{array}{ccc} 1 & kp & p \\ kp & 
k^2re^{i\alpha} & kre^{i\alpha} \\ p & kre^{i\alpha} & re^{i\alpha} \end{array}\right)
\label{mnucol0}
\end{eqnarray}
where 
\begin{equation}    
q^2e^{2i(\beta-\theta)}+p^2 = re^{i\alpha}. 
\label{para1}
\end{equation}
\noindent
Here $m_0$, $k$, $p$, $r$ all are real positive parameters.
We construct the matrix $h(=m_\nu {m_\nu}^{\dagger})$  
to calculate the mixing angles and mass eigenvalues. Expression of $h$ is obtained as 
\begin{equation}
h=m_{\nu}m_{\nu}^{\dagger}= m_{0}^2\begin{pmatrix}
 A &  k\arrowvert B\arrowvert e^{-i\phi} & \arrowvert B\arrowvert e^{-i\phi}  \cr \arrowvert B\arrowvert e^{i\phi} &  k^{2}C & kC
  \cr \arrowvert B\arrowvert e^{i\phi}  &  kC &  C\cr
\end{pmatrix}\label{h}
\end{equation}
where
\begin{eqnarray}
&& A=1+k^{2}p^{2}+p^2 \nonumber\\
&&B= \arrowvert B\arrowvert e^{i\phi}=p+k^{2}pre^{i\alpha}+pre^{i\alpha} \nonumber\\
&&C=p^2+k^{2}r^{2}+r^{2}\nonumber\\
&&\tan \phi =\frac{r\sin \alpha(1+k^2)}{1+r\cos \alpha(1+k^2)} .\label{abccol0}
\end{eqnarray}
 Again factoring out the phase in $h$ as  $h\rightarrow$ $diag(e^{i\phi} ,1,1)$ $h$ $diag(e^{-i\phi} ,1,1)$, finally, we obtain 
\begin{equation}
 h=m_{0}^2A\begin{pmatrix}
 1 &  k\arrowvert B^{\prime}\arrowvert  & \arrowvert B^{\prime}\arrowvert   \cr k\arrowvert B^{\prime}\arrowvert  &  k^{2}C^{\prime}
 & kC^{\prime}  \cr \arrowvert B^{\prime}\arrowvert   &  
kC^{\prime} &  C^{\prime}\cr
\end{pmatrix}
\label{finlh}
\end{equation}
 where $\arrowvert B^{\prime}\arrowvert =
\frac{\arrowvert B\arrowvert }{A}$ and $C^{\prime}=\frac{C}{A}.$
\subsection{Diagonalization}
\begin{figure}
\includegraphics[width=6cm,height=6cm,angle=0]{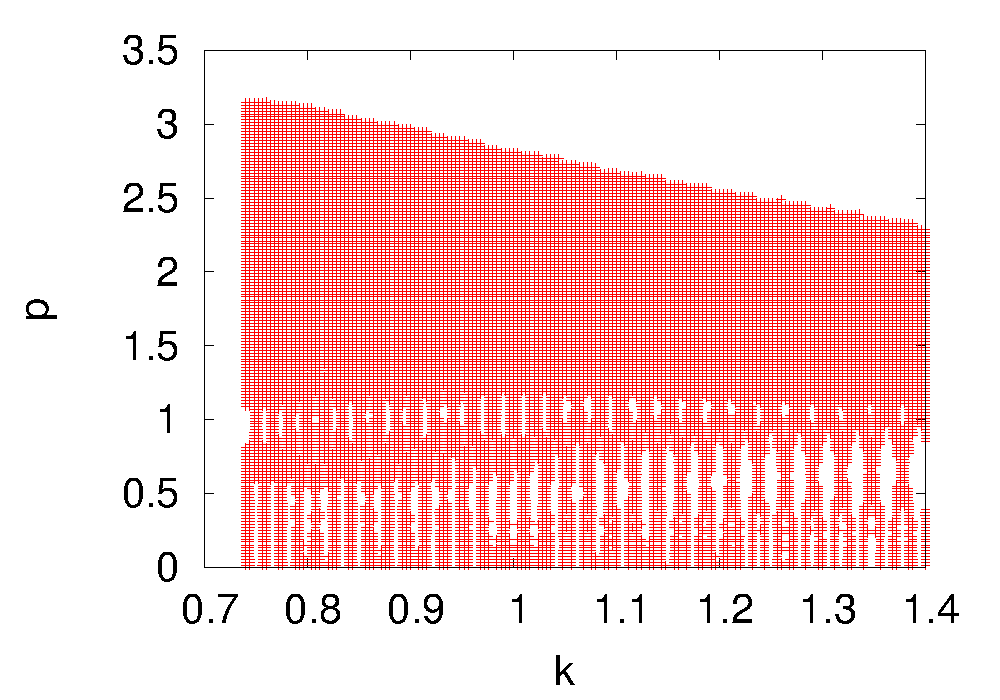}
\includegraphics[width=6cm,height=6cm,angle=0]{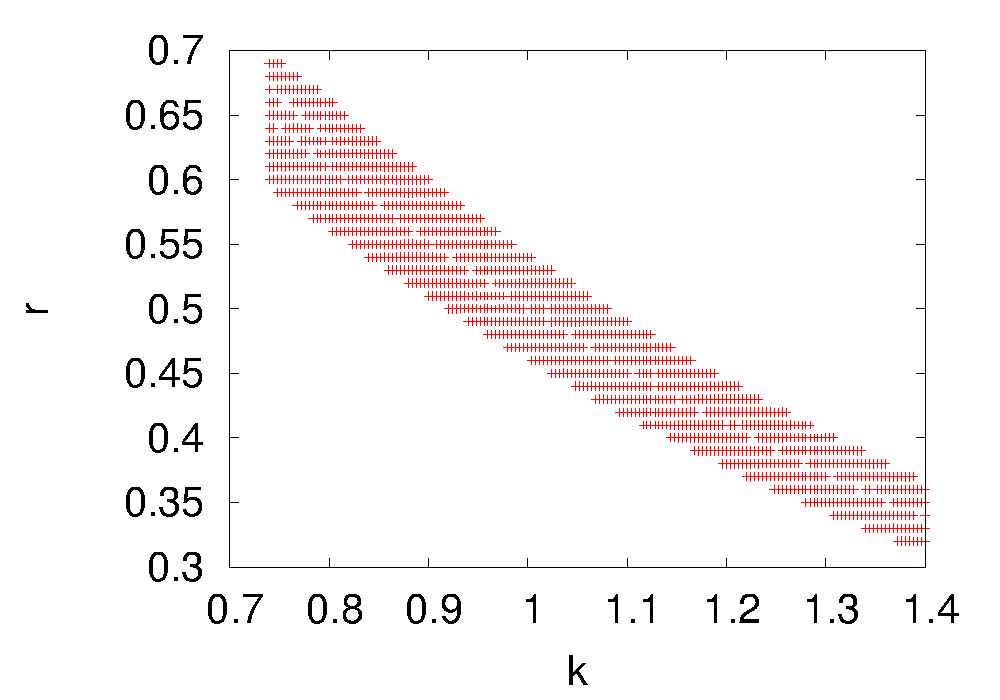}
\caption{Allowed parameter space for ansatz conserving four zero cases 
given in Category A}
\label{unp}
\end{figure}

\vskip 0.1in
\noindent
Diagonalizing the matrix $h$ given in eqn.(\ref{finlh})  
as $U^\dagger h U={\rm diag}(m_1^2,~m_2^2,~m_3^2)$ we get
\begin{eqnarray}
 &&m_{1}^{2}=m^{2}_0A(\frac{P_{1}-\sqrt{P_{1}^{2}-4Q_{1}}}{2})\nonumber\\
 &&m_{2}^{2}=m^{2}_0A(\frac{P_{1}+\sqrt{P_{1}^{2}-4Q_{1}}}{2})\nonumber\\
 &&m_{3}^{2}=0 \label{ev}
\end{eqnarray}
where
\begin{equation}
P_{1}=1+C^{\prime}(k^{2}+1),
Q_{1}=(k^{2}+1)(C^{\prime}-{\arrowvert B ^{\prime}\arrowvert}^{2}),
\label{new1}
\end{equation}
and the mixing matrix is
\begin{equation}
U= \begin{pmatrix}c_{12}&
                      s_{12}&
                        0\cr
-s_{12}c_{23} &c_{12}c_{23}&
s_{23}\cr
s_{12}s_{23} &-c_{12}s_{23} &
c_{23}\cr
\end{pmatrix}
\label{u}
\end{equation}
where $c_{ij}=\cos\theta_{ij}$ and $s_{ij}=\sin\theta_{ij}$.
The three mixing angles are 
\begin{eqnarray}
&&\tan\theta_{23}=-\frac{1}{k} \nonumber\\
 &&\tan\theta_{12}=\frac{2\arrowvert B^{\prime} \arrowvert \sqrt{1+k^{2}}}{C^{\prime}(1+k^{2})-1} \nonumber\\
 &&\theta_{13}=0 \label{t13}
\end{eqnarray}
and the mass squared differences are
\begin{eqnarray}
&&\Delta m_{21}^2= m^{2}_0A\sqrt{P_{1}^{2}-4Q_{1}}\nonumber\\
&&\Delta m_{32}^2=-m^{2}_0A(\frac{P_{1}+\sqrt{P_{1}^{2}-4Q_{1}}}{2}).
\label{delm}
\end{eqnarray}
In Fig.\ref{unp}, we plot the parameter space varying another model parameter 
$\alpha$ within the range  
$-\pi<\alpha<\pi$ satisfying the following $3\sigma$ experimental ranges of 
neutrino data \cite{rslts1,rslts2,rslts3} 
\begin{eqnarray}
&&35.5^\circ\leq \theta_{23} \leq 53.5^\circ\nonumber\\
&&31.7^\circ\leq \theta_{12}\leq 37.7^\circ\nonumber\\
&&6.90\times10^{-5} eV^2\leq (\Delta m^{2}_{21})\leq 8.20\times10^{-5}eV^2\nonumber\\
&&-2.73\times10^{-3}eV^2\leq (\Delta m^{2}_{32}) \leq -1.99\times10^{-3}eV^2.  
\label{3sigmadata}
\end{eqnarray}
We have also used cosmological bound on the sum of the neutrino masses
as  $\Sigma m_{i}<0.5eV$\cite{sm1,sm2,sm3}, and the lower bound obtained
from neutrinoless double beta decay ($\beta\beta_{0\nu}$)
as ${m_{\nu}}_{\beta\beta}<0.35eV$\cite{mnubb}.
\section{Breaking of Scaling ansatz and generation of nonzero $\theta_{13}$}
We want to break the scaling ansatz in such a way that\\
\textbullet \,\, $\theta_{13}$ becomes nonzero.\\
\textbullet \,\,Four zero structure is also retained. \\
\\The second assumption rules out all Category B textures as we have mentioned 
earlier.    
Breaking of scaling ansatz can only be incorporated in the 
remaining six four zero textures in Category A 
and after breaking the scaling ansatz 
by a dimensionless real parameter $\epsilon$ their structure come out 
as follows  
\begin{eqnarray}
&& m_D^{IA}=\left( \begin{array}{ccc} 0 & a & 0 \\ 0 & kb(1+\epsilon) & kc 
\\ 0 & b & c \end{array}\right),
m_D^{IIA} =
\left( \begin{array}{ccc} 0 & 0 & a \\ 0 & kb & kc(1+\epsilon) 
\\ 0 & b & c \end{array}\right),
m_D^{IIIA} = 
\left( \begin{array}{ccc} a & 0 & 0 \\ kb(1+\epsilon) & 0 & kc 
\\ b & 0 & c \end{array}\right),\nonumber\\
&& m_D^{IVA} = \left( \begin{array}{ccc} 0 & 0 & a \\ 
kb & 0 & kc(1+\epsilon) \\ b & 0 & c \end{array}\right),
m_D^{VA} = 
\left( \begin{array}{ccc} a & 0 & 0 \\ kb(1+\epsilon) & kc & 0 \\ 
b & c & 0 \end{array}\right),
m_D^{VIA} = 
\left( \begin{array}{ccc} 0 & a & 0 \\ kb & kc(1+\epsilon) & 0 \\ b & c & 0 \end{array}\right).\nonumber\\
\label{mdsb}
\end{eqnarray}  
Theses structures of $m_D$ are free from RG effects which we have 
discussed in Appendix-A. Moreover, the breaking considered here are the 
most general which can be understood as follows:  
Consider the matrix $m_D^{IA}$ in which the breaking scheme is 
incorporated as
\begin{equation}
{(m_D)}_{\mu 2} = k(1+\epsilon){(m_D)}_{\tau 2}
\label{break1}
\end{equation}
\noindent
while
\begin{equation}
{(m_D)}_{\mu 3} = k{(m_D)}_{\tau 3}.
\label{break2}
\end{equation}
\noindent
Now, redefining the parameters $k(1+\epsilon)\rightarrow k$ and 
$-\epsilon \rightarrow \epsilon$ it is equivalent to break the
ansatz in ${(m_D)}_{\mu 3}$ and ${(m_D)}_{\tau 3}$ elements. Proof 
of this equivalence is similar for other remaining five $m_D$ 
matrices.
\vskip 0.1in
\noindent
The effective neutrino mass matrix $m_{\nu}$ is same for all of 
them and is given by
\begin{equation}
 m_{\nu}=m_0\left( \begin{array}{ccc} 1 & kp+kp\epsilon & p \\ kp+kp\epsilon & 
k^2re^{i\alpha}+2k^2p^2\epsilon & kre^{i\alpha}+kp^2\epsilon \\ p & kre^{i\alpha}+kp^2\epsilon & re^{i\alpha} \end{array}\right)
\end{equation}
with the same definitions of the parameters 
($k$, $p$, $r$, $\alpha$) that we have already used in eqns.(\ref{def}) and 
(\ref{para1}).\\
We now  rewrite this $m_{\nu}$ by  breaking it in two parts, 
one $\epsilon$ dependent and the other independent of $\epsilon$, i.e
\begin{equation}
m_{\nu}=m_0\left( \begin{array}{ccc} 1 & kp & p \\ kp & 
k^2re^{i\alpha} & kre^{i\alpha} \\ p & kre^{i\alpha} & re^{i\alpha} \end{array}\right)
+\epsilon m_{0} \left( \begin{array}{ccc} 0 & kp & 0 \\ kp & 2k^{2}p^{2} & kp^{2}\\
                       0 & kp^{2} & 0 \end{array}\right)  = m_{\nu}^0 + \epsilon m_{\nu}^\prime
\end{equation}
where we have  denoted the first matrix in the right hand side of the 
above equation by $m_\nu^{0}$ and the second one by $m_\nu^\prime$.
Computing  $h_t$ using the above $m_{\nu}$, we get
\begin{equation}
h_{t}=m_{\nu}m_{\nu}^{\dagger}=m_{\nu}^{0}{m_{\nu}^{0}}^{\dagger}
+\epsilon(m_{\nu}^{0}{m_{\nu}^{\prime}}^{\dagger} +m_{\nu}^{\prime}{m_{\nu}^{0}}^{\dagger} )
=h^0+\epsilon h^p
\end{equation}
neglecting O($\epsilon^{2})$ terms. It is to be noted that 
$h^{0}$ is same as $h$, that we have obtained in eq.(\ref{h}). 
After rotating out the phase $\phi$ appearing in $h^{0}$ we are
left with
\begin{equation}
h_{t}^\prime=m_{0}^2\begin{pmatrix}
 A &  k\arrowvert B\arrowvert  & \arrowvert B\arrowvert   \cr k\arrowvert B\arrowvert  &  k^{2}C & kC
  \cr \arrowvert B\arrowvert   &  kC &  C\cr
\end{pmatrix}+ \epsilon h^{\prime\prime} \label{hp}
\end{equation}
where $h^{\prime\prime}=diag(e^{i\phi} ,1,1)$ $h^{p}$ $diag(e^{-i\phi} ,1,1)$ and
$h_t^\prime=diag(e^{i\phi} ,1,1)$ $h_t$ $diag(e^{-i\phi} ,1,1)$. To diagonalise $h_{t}^\prime$
 we first rotate this matrix with unperturbed diagonalising matrix $U$ in eq.\ (\ref{u}) with angles in eq.\ (\ref{t13}).
 The first part of $h_t^\prime$ becomes diagonal, however, the $h^{\prime\prime}$ part is not. Performing the operation 
$ U^{\dagger}h_{t}^\prime U$ we get
\begin{equation}
 {h_{t}}^{\prime\prime}=U^{\dagger}h_{t}^\prime 
U=\begin{pmatrix}m_{1}^{2} & 0 & 0 \cr 0 & m_{2}^{2} & 0 \cr 0 &  0 & 0 \cr
\end{pmatrix}
+\epsilon \begin{pmatrix}x & y & z \cr y^{\ast} & w & q \cr z^{\ast} &  
q^{\ast} & 0 \cr
\end{pmatrix} 
\label{htp}
\end{equation}
where different elements of the the 2nd matrix are obtained from the explicit multiplication 
$U^{\dagger}h^{\prime\prime}U$.
To diagonalise the second matrix of $h_{t}^{\prime\prime}$ we further require the matrix 
\begin{equation}
U_{\epsilon}=\begin{pmatrix}1 & \epsilon a & \epsilon b
\cr -\epsilon a^{\ast} & 1 & \epsilon c \cr -\epsilon b^{\ast} & -
\epsilon c^{\ast} & 1\cr\end{pmatrix}.
\end{equation}
\noindent
Explicit expressions of parameters $x$, $y$, $z$, $q$ and $w$ are 
given in Appendix B. 
We demand that upto O($\epsilon$)the above matrix diagonalises  $ h_{t}^{\prime\prime}$ of eq.(\ref{htp}), i.e after 
the operation
${U_{\epsilon}}^{\dagger}{h_{t}}^{\prime\prime}U_{\epsilon}$ 
the off-diagonal elements of the resulting matrix are zero
 and solving those equations we find out the unknown variables 
$a$, $b$, $c$. They come out as 
\begin{eqnarray}
&&a=\frac{y}{(m_{2}^{2}-m_{1}^{2})}\nonumber\\
&&b=-\frac{z}{m_{1}^{2}}\nonumber\\
&&c=-\frac{q}{m_{2}^{2}} .
\end{eqnarray}
As a result of this rotation by the matrix $U_{\epsilon}$ we get 
\begin{equation}
 U_{\epsilon}^{\dagger}h_{t}^{\prime\prime}U_{\epsilon}=\begin{pmatrix}
m_{1}^{2}+\epsilon x & 0
& 0\cr 0 & m_{2}^{2}+\epsilon w & 0\cr 0 & 0 & 0\cr
\end{pmatrix}.
\label{mas}
\end{equation}
In a concise way, we actually have done the following  
\begin{equation}
 U_{\epsilon}^{\dagger}U^{\dagger}h_{t}^\prime UU_{\epsilon}=
\begin{pmatrix} m_{1}^{2}+\epsilon x & 0
& 0\cr 0 & m_{2}^{2}+\epsilon w & 0\cr 0 & 0 & 0 \cr
\end{pmatrix}
=\begin{pmatrix} {m_{1}^{\prime}}^{2} & 0
& 0\cr 0 & {m_{2}^{\prime}}^{2} & 0\cr 0 & 0 & {m_3^\prime}^2 \cr
\end{pmatrix}
\end{equation}
where $m_{1}^{\prime}$, $m_{2}^{\prime}$, $m_{3}^{\prime}$ are the new mass eigenvalues. $m_3^\prime$ is
still zero even after breaking of scaling ansatz because one column remain zero for all allowed $m_D$.
 Hence, the total diagonalisation matrix in our scheme is  $ V=UU_{\epsilon}$.
Explicitly $V$ is given by
\begin{eqnarray}
\small{V=\begin{pmatrix} c_{12}+s_{12}
(\frac{\epsilon y^{\ast}}{m_{1}^{2}-m_{2}^{2}}) & s_{12}+c_{12}
(\frac{\epsilon y}{m_{2}^{2}-m_{1}^{2}})
& -c_{12}(\frac{\epsilon z}{m_{1}^{2}})-s_{12}(
\frac{\epsilon q}{m_{2}^{2}})\nonumber \cr 
&&\nonumber\cr
-c_{23}s_{12}+
c_{12}c_{23}
(\frac{\epsilon y^{\ast}}{m_{1}^{2}-m_{2}^{2}})
&-c_{23}s_{12}(\frac{\epsilon y}{m_{2}^{2}-m_{1}^{2}})+c_{12}c_{23}
& c_{23}s_{12}
(\frac{\epsilon z}{m_{1}^{2}})-c_{12}c_{23}
(\frac{\epsilon q}{m_{2}^{2}})
\nonumber\cr
+s_{23}
(\frac{\epsilon z^{\ast}}{m_{1}^{2}}) 
& +s_{23}(
\frac{\epsilon q^{\ast}}{m_{2}^{2}}) &
+s_{23}
\nonumber \cr 
&&\nonumber\cr
s_{23}s_{12}-s_{23}c_{12}(
\frac{\epsilon y^{\ast}}{m_{1}^{2}-m_{2}^{2}})
&
s_{23}s_{12}(\frac{\epsilon y}{m_{2}^{2}-m_{1}^{2}})-s_{23}c_{12}
&
-s_{23}s_{12}
(\frac{\epsilon z}{m_{1}^{2}})
+s_{23}c_{12}(\frac{\epsilon q}{m_{2}^{2}})
\nonumber\cr
+c_{23}
(\frac{\epsilon z^{\ast}}{m_{1}^{2}}) 
& 
+c_{23}
(\frac{\epsilon q^{\ast}}{m_{2}^{2}}) 
& 
+c_{23}\cr.
\end{pmatrix}}
\nonumber\\
\end{eqnarray}
To find out the three mixing angles we have to compare $V$ with PMNS matrix. The $U_{\rm PMNS}$ is given by
\begin{equation}
U_{\rm PMNS}= \begin{pmatrix}c_{12}^\prime c_{13}^\prime&
                      s_{12}^\prime c_{13}^\prime&
                      s_{13}^\prime e^{-i\delta}\cr
-s_{12}^\prime c_{23}^\prime-c_{12}^\prime s_{23}^\prime s_{13}^\prime e^{i\delta}&c_{12}^\prime c_{23}^\prime-
s_{12}^\prime s_{23}^\prime s_{13}^\prime e^{i\delta}&
s_{23}^\prime c_{13}^\prime\cr
s_{12}^\prime s_{23}^\prime -c_{12}^\prime c_{23}^\prime s_{13}^\prime e^{i\delta}&
-c_{12}^\prime s_{23}^\prime -s_{12}^\prime c_{23}^\prime s_{13}^\prime e^{i\delta}&
c_{23}^\prime c_{13}^\prime\cr
\end{pmatrix}
\begin{pmatrix}e^{i\alpha_M}&0&0\cr
         0&e^{i\beta_M}&0\cr
         0&0&1,
\end{pmatrix}
\end{equation}
(with $c_{ij}^\prime = cos\,\theta_{ij}^\prime$, $s_{ij}^\prime = sin\,\theta_{ij}^\prime$, 
$\delta$ is the Dirac phase and $\alpha_M$, $\beta_M$ are the Majorana phases.)\\
After neglecting the higher 
order terms in
$\epsilon$ the modified mixing angles are given by
\begin{eqnarray}
&&\tan \theta_{23}^{\prime}=\frac{\left|V_{23}\right|}{\left|V_{33}\right|}\approx t_{23}+\epsilon (1+t_{23}^{2})(\frac{s_{12}}{m_{1}^{2}}Re(z)-\frac{c_{12}}{m_{2}^{2}}Re(q))\nonumber\\
&&\tan \theta_{12}^{\prime}=\frac{\left|V_{12}\right|}{\left|V_{11}\right|}\approx  t_{12}+\epsilon (1+t_{12}^{2})\frac{Re(y)}{(m_{2}^{2}-m_{1}^{2})}\nonumber\\
&&\sin \theta_{13}^{\prime}=\left|V_{13}\right|\approx \epsilon \sqrt{(\frac{c_{12}}{m_{1}^{2}}Re(z)+\frac{s_{12}}{m_{2}^{2}}Re(q))^{2}+
     (\frac{c_{12}}{m_{1}^{2}}Im(z)+\frac{s_{12}}{m_{2}^{2}}Im(q))^{2}}\nonumber\\
\end{eqnarray}
and the CP violating phase $\delta$ is given by
\begin{equation}
\tan \delta =\frac{\frac{c_{12}}{m_{1}^{2}}Im(z)+\frac{s_{12}}{m_{2}^{2}}Im(q)}{\frac{c_{12}}{m_{1}^{2}}Re(z)+\frac{s_{12}}{m_{2}^{2}}Re(q)}
\end{equation}
where $c_{ij} = \cos\theta_{ij}$, $s_{ij} = \sin\theta_{ij}$, $t_{ij} = 
\tan\theta_{ij}$  are the mixing angles and $m_1$, $m_2$ are the masses for the ansatz conserving 
case of eqn.\ (\ref{t13}) and eqn.\ (\ref{ev}) respectively. 
From eqn.\ (\ref{mas}) we have the mass squared differences:
\begin{eqnarray}
&&(\Delta m_{21}^2)^\prime= \Delta m_{21}^2 +\epsilon(w-x)\nonumber\\
&&(\Delta m_{32}^2 )^\prime=\Delta m_{32}^2 - \epsilon w
\end{eqnarray}
where $\Delta m_{21}^2$ and $\Delta m_{32}^2$ are mass squared differences for unperturbed  scaling ansatz as in 
eq.\ (\ref{delm}).
The measure of CP violation is understood through $J_{CP}$ which is defined as
\begin{eqnarray}
 J_{\rm CP}= \frac{(h_t)_{12} (h_t)_{23} (h_t)_{31}}{
(\Delta m_{21}^2)^\prime (\Delta m_{32}^2)^\prime (\Delta m_{31}^2)^\prime} 
\end{eqnarray}
which is known function of $k$, $p$, $r$, $\alpha$ and $\epsilon$.
\section{Discussion of numerical results}
We explore the parameter space of the above case using the same 
$3\sigma$ values of neutrino experimental data given in 
eqn.(\ref{3sigmadata}).  
The Lagrangian parameters $p$ and $r$
are ranging from zero to some positive values 
since we have separated out the phase part from them.
The scale factor $k$ should not have zero value because in this
case the second row of $(m_D)$ is zero which in turn decouple the 
second generation.     
The constrained parameter space we obtain as 
\begin{eqnarray}
 &&0<p<4\nonumber\\&&0.65<k<1.4\nonumber\\&&0<r<0.7
\nonumber\\&&-180^\circ< \alpha < 180^\circ .\label{prange}
\end{eqnarray}
\begin{figure}
\includegraphics[width=6cm,height=6cm,angle=0]{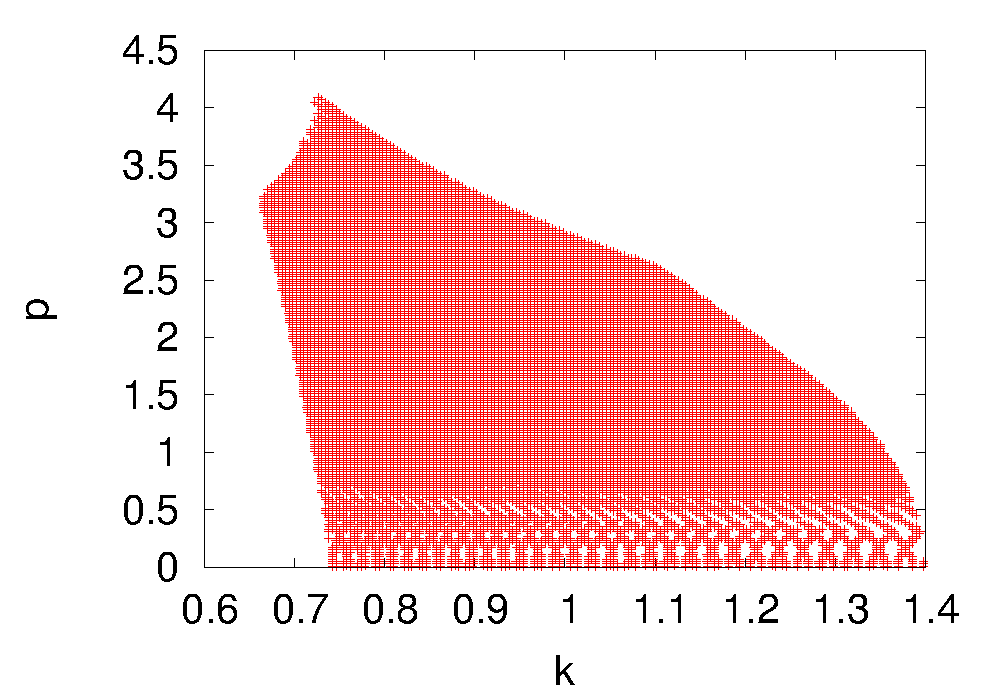}
\includegraphics[width=6cm,height=6cm,angle=0]{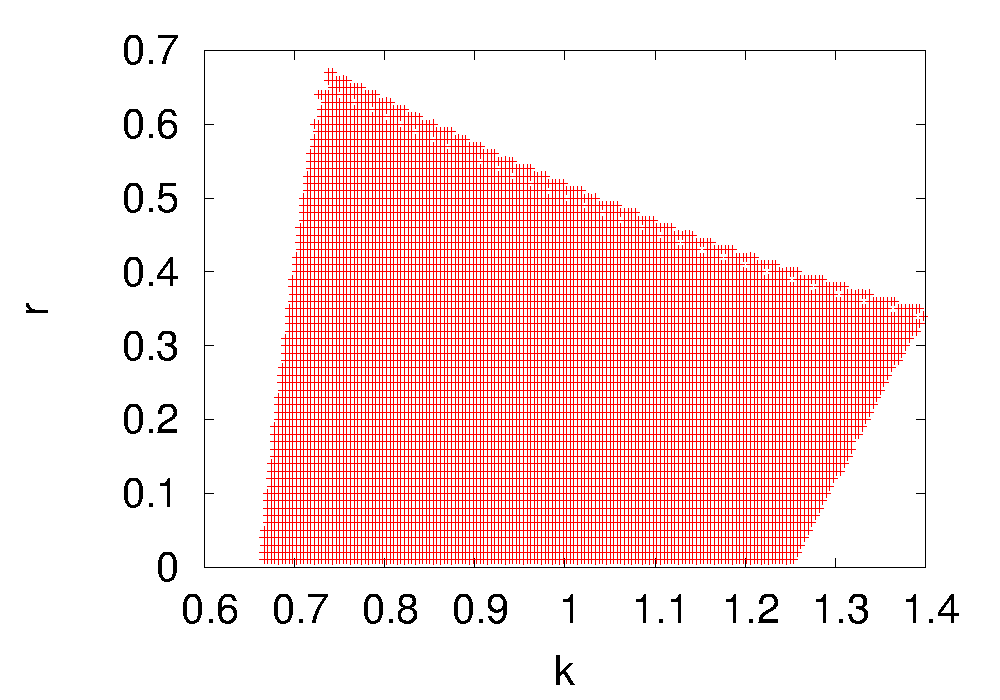}
\caption{Allowed parameter space for $\epsilon=0.1$}
\label{1}
\end{figure}
The values outside this range is not 
admissible within the above mentioned experimental ranges. 
It is to be noted that allowed parameter space in $k$-$r$ plane  
for the ansatz breaking case is much larger than that
in the ansatz conserving case. First of all, we found that
throughout the allowed parameter space $\Sigma m_{i}$ 
and ${m_{\nu}}_{\beta\beta}$ are always far below the experimental
bounds which could be hardly tested in the near future experiments.
Next, it is amply clear from the expression of 
$\theta_{13}$ that it is directly proportional to the value of the 
ansatz breaking parameter $\epsilon$. The parameter $\epsilon$ is varied upto a
reasonable choice $\epsilon=0.1$ for which a large $\theta_{13}$ is generated, 
however,
for a smaller value of $\epsilon$ such as $\epsilon \sim 0.07$, $\theta_{13}\sim 10^\circ$ is also 
admitted because present experimental bound on $\theta_{13}$ is $3.75^\circ\le\theta_{13}\le 13.60^\circ$ for $3\sigma$ 
 bound from RENO \cite{last} and $4.90^\circ\le\theta_{13}\le 11.51^\circ$ for $3\sigma$ 
 bound from Daya-Bay \cite{DayaBay}. We have shown all plots for a representative value of $\epsilon=0.1$ . The allowed Lagrangian 
parameter space is plotted in Fig.\ref{1}.
\begin{figure}
\includegraphics[width=6cm,height=6cm,angle=0]{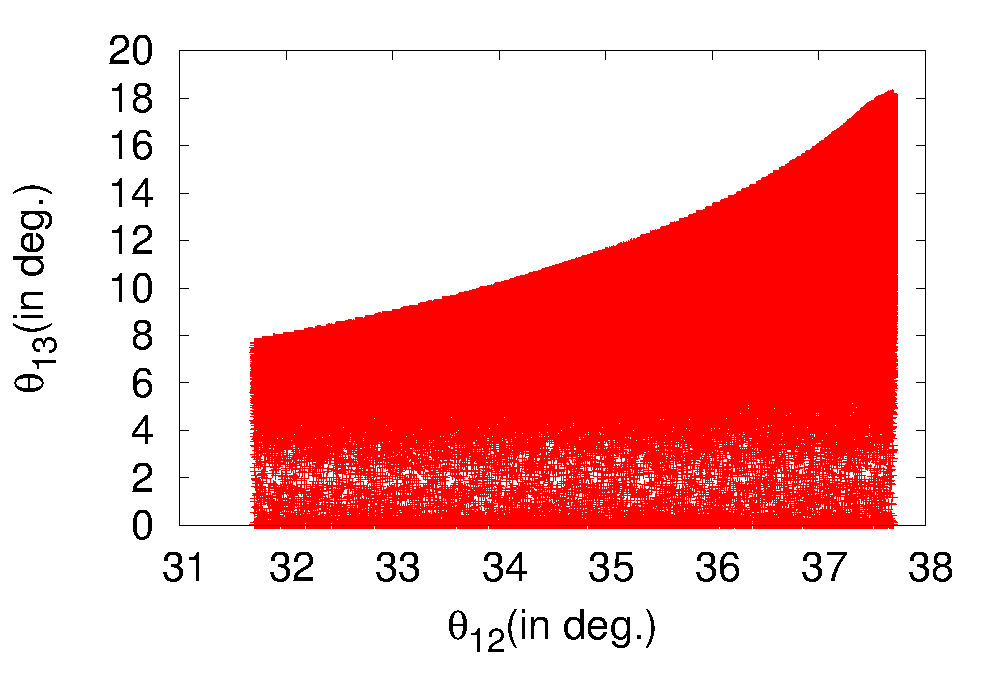}
\includegraphics[width=6cm,height=6cm,angle=0]{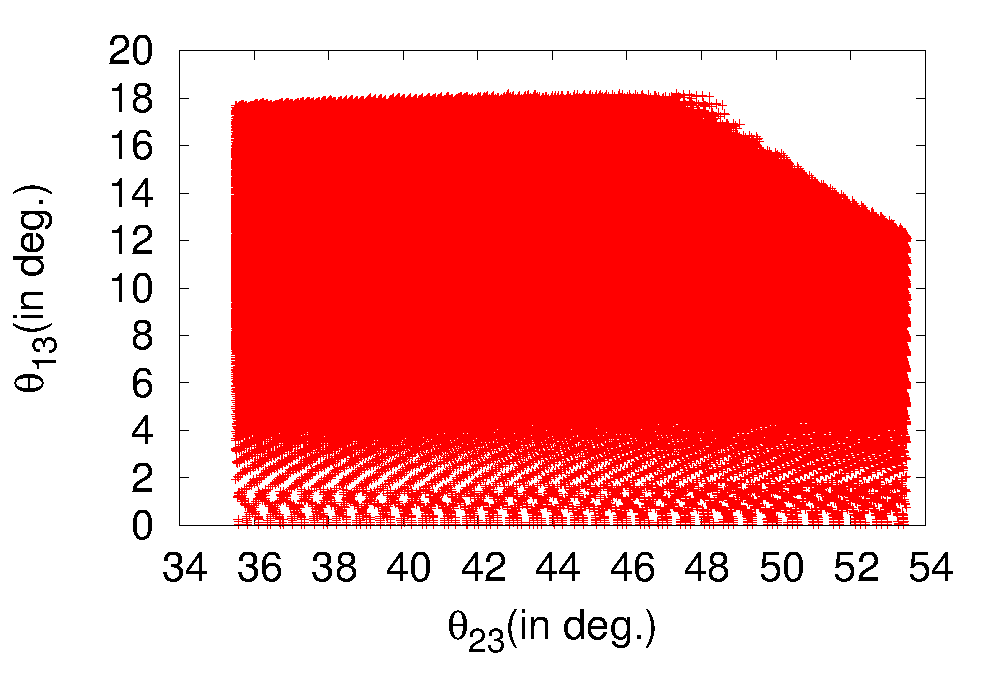}
\caption{Allowed values of the mixing angles for $\epsilon=0.1$}
\label{2}
\end{figure}
From Fig.\ref{2} it is clear 
that $\theta_{13}$ is almost insensitive to $\theta_{23}$, however, 
 significantly related to the values of $\theta_{12}$ 
which is depicted in Fig.3.
The CP violation parameter $J_{cp}$ arises due to nonzero $\theta_{13}$ is plotted with $\theta_{13}$ in Fig.\ref{3}. 
Sign of $\delta$ is not constrained from oscillation experiments it needs further calculation of baryon asymmetry.
Plot (Fig.\ref{3}) of $\theta_{13}$ vs $\delta$ shows that $\delta$ is maximum for smaller values of $\theta_{13}$ and 
for larger values of $\theta_{13}$, $\delta$ is relatively small. If we restrict $\theta_{13}$ in $3\sigma$
experimental range we have the bound on $\delta$, $0\le\delta\le 35^\circ$ and $J_{CP}$, $0\le J_{CP}\le 0.02$.
\begin{figure}
\includegraphics[width=6cm,height=6cm,angle=0]{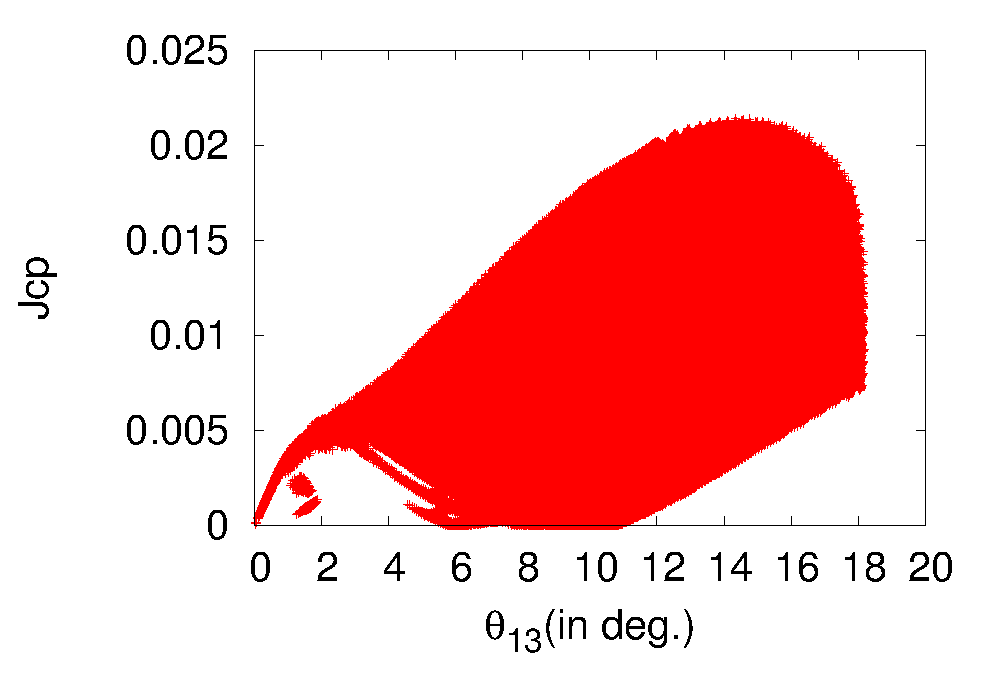}
\includegraphics[width=6cm,height=6cm,angle=0]{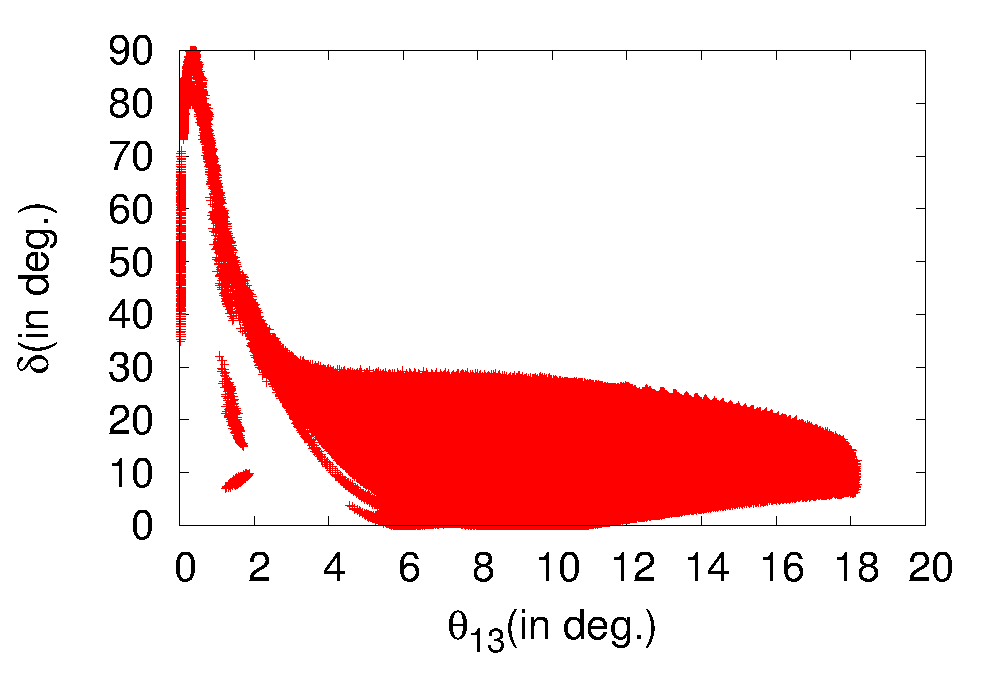}
\caption{Allowed $\arrowvert \ J_{CP} \arrowvert$ vs $\theta_{13}$ (left), 
  $\arrowvert \delta \arrowvert$ vs $\theta_{13}$(right) for $\epsilon=0.1$  }
\label{3}
\end{figure}
\section{Leptogenesis with broken scaling ansatz}
\subsection{General discussion on Leptogenesis and Baryogenesis}
Let us briefly discuss about right handed Majorana neutrino decay
generated leptogenesis. There is a Dirac type Yukawa interaction 
of right handed neutrino (${N}_i$) with SM lepton
doublet and Higgs doublet. At the energy scale where $SU(2)_L\times
U(1)_Y$ symmetry is preserved, physical right handed neutrino ${N}_i$
with definite mass can decay both to charged lepton with charged
scalar and light neutrino with neutral scalar. Due to the Majorana character
of ${N}_i$, conjugate process is also possible. If out of equilibrium
decay of ${N}_i$ in conjugate process occur at different rate from
actual process, net lepton number will be generated. The CP asymmetry
of decay is characterized by a parameter $\varepsilon_i$ which is
defined as
\begin{eqnarray}
\varepsilon_i &=&\frac{\Gamma_{{N}_i\rightarrow
    l^-\phi^+,\nu_l\phi^0}-\Gamma_{{N}_i\rightarrow
    l^+\phi^-,\nu_l^c\phi^{0*}}}{\Gamma_{{N}_i\rightarrow
    l^-\phi^+,\nu_l\phi^0}+\Gamma_{{N}_i\rightarrow
    l^+\phi^-,\nu_l^c\phi^{0*}}}.
\label{cpasym}
\end{eqnarray}
We are working in a basis where right handed neutrinos have definite mass as , 
$M_R={\rm diag}(M_1, M_2, M_3)$. Now, the decay
asymmetry $\varepsilon_i$ for $N_i$ decay occurs at one loop level. Interference of tree level, one loop vertex and 
self energy diagrams generate the following $\varepsilon_i$ for hierarchical right handed neutrino mass spectrum: 
\begin{eqnarray}
\varepsilon_i=\frac{1}{4\pi v^2H_{ii}}\sum_{j\ne
  i}Im(H_{ij}^2)f(x_{ij})
\label{asy}
\end{eqnarray}
where $x_{ij}=M_j^2/M_i^2$, $H=m_D^\dagger m_D$ and \cite{Covi:1996wh}
\begin{eqnarray}
 f(x)=\sqrt{x}\left\{1-(1+x)\ln(1+\frac{1}{x})+\frac{1}{1-x}\right\}.
\end{eqnarray}
CP asymmetry parameters $\varepsilon_i$ are related
to the leptonic asymmetry parameters through $Y_L$ as
\cite{Nielsen:2001fy,Pilaftsis:2003gt,Antusch:2006cw}
\begin{eqnarray}
Y_L\equiv\frac{n_L-{\bar n}_L}{s}=\sum_i^3\frac{\varepsilon_i\kappa_i}{g_{*i}}
\label{leptasym}
\end{eqnarray}
where $n_L$ is the lepton number density, ${\bar n}_L$ is the
anti-lepton number density, $s$ is the entropy density, $\kappa_i$ is
the dilution factor for the CP asymmetry $\varepsilon_i$ and $g_{*i}$
is the effective number of degrees of freedom \cite{Roos:1994fz} at
temperature $T=M_i$. The baryon asymmetry $Y_B$
produced through the sphaleron transmutation of $Y_L$, while the
quantum number $B-L$ remains conserved, is given by
\cite{Harvey:1990qw}
\begin{eqnarray}
Y_B=-\frac{8N_F+4N_H}{22N_F+13N_H}Y_L 
\label{barasym}
\end{eqnarray}
where $N_F$ is the number of fermion families and $N_H$ is the number
of Higgs doublets. The quantity $Y_B=-\frac{28}{79} Y_L$ in eq.\ (\ref{barasym})
for SM. Now we introduce the relation between $Y_B$ and $\eta_B$,
where $\eta_B$ is the baryon number density over photon number density
$n_\gamma$. The relation is \cite{Kolb}
\begin{eqnarray}
\eta_B=\left.\frac{s}{n_\gamma}\right|_0Y_B=7.0394Y_B,
\label{yetar}
\end{eqnarray}
where the zero indicates present time. Finally we have relation between $\eta_B$ and $\varepsilon_i$
\begin{eqnarray}
\eta_B=-2.495\times\sum_i\frac{\varepsilon_i\kappa_i}{g_{*i}}.
\label{etab}
\end{eqnarray}
This dilution factor $\kappa_i$ approximately given by \cite{Giudice:2003jh, Buchmuller:2004nz, Abada:2006ea}
\begin{eqnarray}
\frac{1}{\kappa_i}\simeq\frac{8.25}{K_i} +\left(\frac{K_i}{0.2}\right)^{1.16}\qquad {\rm with}\qquad  K_i=\frac{\Gamma_i}{H_i},
\label{kppa}
\end{eqnarray}
where $\Gamma_i$ is the decay width of $N_i$ and $H_i$ is Hubble
constant at $T=M_i$. Their expressions are
\begin{eqnarray}
\Gamma_i=\frac{h_{ii}M_i}{4\pi v^2}\qquad {\rm and} \qquad H_i=1.66\sqrt{g_{*i}}\frac{M_i^2}{M_P},
\label{gh}
\end{eqnarray}
where $v=246$GeV and $M_P=1.22\times 10^{19}$GeV. Thus we have 
\begin{eqnarray}
K_i=\frac{M_PH_{ii}}{1.66\times 4\pi\sqrt{g_{*i}}v^2M_i}.
\label{K1}
\end{eqnarray}
\subsection{Calculation of lepton and baryon asymmetry with broken scaling ansatz}
The matrix $H=m_D^\dagger m_D$ shown in eq.\ (\ref{asy}) 
is important to study leptogenesis. 
For six possible $m_D$ with broken scaling ansatz by
parameter $\epsilon$ are given in eq.\ \ref{mdsb}. 
They will generate following six possible $H$ in three pairs:
\begin{eqnarray}
&& m_0M_2\left( \begin{array}{ccc} 0 & 0 & 0 \\ 0 & 1+p^2(1+k^2)+2p^2k^2\epsilon & lpqe^{i(\beta -\theta)}(1+k^2+k^2\epsilon) 
\\ 0 & lpqe^{-i(\beta -\theta)}(1+k^2+k^2\epsilon)  &  q^2l^2(1+k^2)\end{array}\right)\quad {\rm with }\quad
l=\sqrt{\frac{M_3}{M_2}}
\nonumber\\\nonumber\\\nonumber\\
&&m_0M_3\left( \begin{array}{ccc} 0 & 0 & 0 \\ 0 &  q^2l^2(1+k^2) & lpqe^{-i(\beta -\theta)}(1+k^2+k^2\epsilon) 
\\ 0 & lpqe^{i(\beta -\theta)}(1+k^2+k^2\epsilon)  &  1+p^2(1+k^2)+2p^2k^2\epsilon\end{array}\right)
\quad {\rm with }\quad l=\sqrt{\frac{M_2}{M_3}}\nonumber\\
\label{p1}
\end{eqnarray}
 \begin{eqnarray}
&& m_0M_1\left( \begin{array}{ccc} 1+p^2(1+k^2)+2p^2k^2\epsilon & 0 & lpqe^{i(\beta -\theta)}(1+k^2+k^2\epsilon) \\ 0 & 0 & 0 
\\ lpqe^{-i(\beta -\theta)}(1+k^2+k^2\epsilon) & 0  &  q^2l^2(1+k^2)\end{array}\right)\quad {\rm with }\quad
l=\sqrt{\frac{M_3}{M_1}}
\nonumber\\\nonumber\\\nonumber\\
&&m_0M_3\left( \begin{array}{ccc} q^2l^2(1+k^2) & 0 & lpqe^{-i(\beta -\theta)}(1+k^2+k^2\epsilon)\\
 0 &  0 & 0
\\ lpqe^{i(\beta -\theta)}(1+k^2+k^2\epsilon) & 0  &  1+p^2(1+k^2)+2p^2k^2\epsilon\end{array}\right)
\quad {\rm with }\quad l=\sqrt{\frac{M_1}{M_3}}\nonumber\\
\end{eqnarray}   
 \begin{eqnarray}
&& m_0M_1\left(\begin{array}{ccc} 1+p^2(1+k^2)+2p^2k^2\epsilon &  lpqe^{i(\beta -\theta)}(1+k^2+k^2\epsilon) & 0
\\ lpqe^{-i(\beta -\theta)}(1+k^2+k^2\epsilon) & q^2l^2(1+k^2) & 0\\0 & 0 & 0  \end{array}\right)
\quad {\rm with }\quad l=\sqrt{\frac{M_2}{M_1}}\nonumber\\\nonumber\\\nonumber\\
&&m_0M_2\left( \begin{array}{ccc} q^2l^2(1+k^2) &  lpqe^{-i(\beta -\theta)}(1+k^2+k^2\epsilon) & 0\\
 lpqe^{i(\beta -\theta)}(1+k^2+k^2\epsilon)   &  1+p^2(1+k^2)+2p^2k^2\epsilon & 0
\\0 & 0 & 0  \end{array}\right)\quad {\rm with }\quad l=\sqrt{\frac{M_1}{M_2}}.\nonumber\\
\end{eqnarray}   
Parameters in above six possible $H$ are already defined in eq.(\ref{def}) 
and only $l$ is defined here along with every $H$.
Interesting features of the three pairs of $H$ are that for every pair one generation of right handed neutrino decouples
and also its decay width vanishes and hence could not take part in generation of lepton asymmetry. For the first pair
$N_1$ decouples, for the 2nd pair $N_2$ decouples and for the 3rd pair $N_3$ decouples. Apart from this one more interesting
point is that first matrix of every pair have similar expression in their non-zero diagonal and off-diagonal elements whereas
the 2nd matrix of every pair have similar expressions. So, we don't need to study all the three pairs. We will
only study the first pair. 

First generation of right handed neutrino $N_1$ decay 
width is zero. Lepton asymmetry is generated through decay
 of $N_2$ and $N_3$ only
contribute. Decay asymmetries $\varepsilon_2$ and $\varepsilon_3$ for the first form of the first pair in 
eq.\ (\ref{p1}),  
 \begin{eqnarray}
&& \varepsilon_2=\frac{1}{4\pi v^2} \frac{{\rm Im}(H_{23}^2)}{H_{22}}f(M_3^2/M_2^2)=\frac{M_2m_0}{4\pi v^2} F f(l^4)\nonumber\\
&&\varepsilon_3=\frac{1}{4\pi v^2} \frac{{\rm Im}(H_{32}^2)}{H_{33}}f(M_2^2/M_3^2)=-\frac{M_2m_0}{4\pi v^2} F' f(1/l^4)
 \end{eqnarray}
where $l=\sqrt{\frac{M_3}{M_2}}$ and 
\begin{eqnarray}
&&F=\frac{rl^2p^2(1+k^2)\sin\alpha}{1+p^2(1+k^2)}\left[1+k^2+\frac{2k^2\epsilon}{1+p^2(1+k^2)}\right]\nonumber\\
&&F'=\frac{rp^2(1+k^2)\sin\alpha}{(1+k^2)\sqrt{p^4+r^2-2rp^2\cos\alpha}}\left[1+k^2+2k^2\epsilon \right].
\end{eqnarray}
The definition of different parameters for different $m_D$ are given in 
eq.\ \ref{def} and also we have used
 $q^2e^{2i(\beta -\theta)}=re^{i\alpha}-p^2$. The washout factors for 2nd and 3rd generation are
\begin{eqnarray}
&&K_2=\frac{M_PH_{22}}{1.66\times 4\pi\sqrt{g_{*2}}v^2M_2}=913.7\left(\frac{m_0}{\rm eV}\right)
\left[1+p^2(1+k^2+2k^2\epsilon)\right]\nonumber\\
&&K_3=\frac{M_PH_{33}}{1.66\times 4\pi\sqrt{g_{*3}}v^2M_3}=913.7\left(\frac{m_0}{\rm eV}\right)
(1+k^2)\sqrt{p^4+r^2-2rp^2\cos\alpha}.\nonumber\\
\label{K}
\end{eqnarray}
where we have used $v=246$GeV, $M_P=1.22\times 10^{19}$GeV and $g_{*i}=110.25$ 
for SM with two right handed neutrinos.
With this washout factors we can determine the dilution factors $\kappa_2$ and $\kappa_3$ using the formula given in 
eq.\ \ref{kppa}. Well equipped with the above formulae for $\varepsilon_2$, $\varepsilon_3$, $\kappa_2$ and $\kappa_3$ we
can easily generate the expression for baryon asymmetry
\begin{eqnarray}
\eta_B&=&-2.495\times\sum_i\frac{\varepsilon_i\kappa_i}{g_{*i}}\nonumber\\
&=&-2.27\times10^{-2}\left[\varepsilon_2\kappa_2+\varepsilon_3\kappa_3\right].
\label{etab1}
\end{eqnarray}
An additional beauty is that the expressions for $\eta_B$ for two matrices in a pair are same. For the 2nd matrix of the 
first
pair in eq.\ (\ref{p1}), expressions for  $\varepsilon_2$ and $K_2$ are same as the expressions of $\varepsilon_3$ and $K_3$ for
 the first matrix of the
pair and expressions for  $\varepsilon_3$ and $K_3$ are same as the expressions of $\varepsilon_2$ and $K_2$ of
 first matrix of the
pair. So, effectively $\eta_B$ expression remains same. Consequence is same as for the first matrix of the pair.

The expression of $\eta_B$ depends on $m_0$, $k$, $p$, $r$, $\alpha$, $\epsilon$ and additional two parameters $M_2$ and 
$l=\sqrt{\frac{M_3}{M_2}}$. On the top of constrained  parameter space from neutrino data, 
we have also explored the parameter space with the additional constraint arises due to 
baryon asymmetry 
$5.5\times 10^{-10}< \eta_b< 7\times 10^{-10}$ \cite{bary1,bary2,bary3} for $0.1\le l\le 0.9$ and 
$1.1\le l\le 10$ (avoiding point of degeneracy $l=1$) and $10^{12}~{\rm GeV}\le M_2\le 10^{15}~{\rm GeV}$. 
We have seen that change in the parameter space is negligible. Only sign of $\alpha$ is constrained for different $l$.
For $1.1\le l\le 1.54$ and $0.1\le l\le 0.9$ sign of $\alpha$ is negative and for $1.54\le l\le 10$ 
sign of $\alpha$ is positive. Again  $\alpha=0,\pm 180$ are not allowed. But value of $\alpha$ near $0$ and $180$ are
still allowed and the $M_2$ value is large there, $M_2\simeq O(10^{15})$ GeV.  
\section{Summary}
To sum up, we have explored a predictive and testable scenario of neutrino 
mass matrix accommodating scaling ansatz with four zero Yukawa 
textures advocating type I seesaw mechanism with diagonal charged leptons 
and right chiral neutrino mass matrices. 
We break scaling ansatz in the Yukawa matrices to generate nonzero $\theta_{13}$ through a dimensionless 
parameter $\epsilon$. 
The parameter space of the textures 
studied   allow the $3\sigma$ value of $\theta_{13}$ along with other neutrino experimental data. 
Using the $\theta_{13}$ constraint we have restricted
Dirac CP phase $\delta$ and $J_{CP}$. We have also studied baryogenesis 
via leptogenesis arises in those textures, however, 
there is no drastic change in the parameter space due to the constraint from baryogenesis. But, the sign 
of the only phase present in this model is fixed.
\\\\
{\bf Acknowledgement}
\\ Authors would like to thanks Probir Roy, Debabrata Adak and Anirban Biswas for many helpful discussions.
\appendix
\section{RG Effect}
It is to be noted that even after breaking of scaling anasatz, the textures 
given in eqn.(\ref{mdsb}) are invariant under RG evolution. 
This is guaranteed in the following way : 
\vskip 0.1in
\noindent
Following the methodology presented in Ref.\cite{rg1}- \cite{rg2}
due to $\tau$ lepton mass correction on $m_D$ of eqn. 
(\ref{mD}) with scaling ansatz we get 
\begin{equation}
\begin{pmatrix}1 & 0 &0\cr
               0& 1 & 0\cr
               0& 0 & 1-\Delta_\tau
\end{pmatrix}
\begin{pmatrix} a_1 & a_2 &a_3\cr
                kb_1& kb_2 & kb_3\cr
                b_1 & b_2 &b_3
\end{pmatrix}
= 
\begin{pmatrix}   a_1 & a_2 &a_3\cr
                  kb_1& kb_2 & kb_3\cr
                  b_1(1-\Delta_\tau) & b_2(1-\Delta_\tau) & b_3(1-\Delta_\tau)

\end{pmatrix}.
\label{rg1}
\end{equation}
\noindent
Redifining $b_i$`s as $b_1(1-\Delta_\tau)\rightarrow b_1$, 
$b_2(1-\Delta_\tau)\rightarrow b_2$, 
$b_3(1-\Delta_\tau)\rightarrow b_3$ 
we  get 
\begin{equation}
m_D = \begin{pmatrix}a_1 & a_2 & a_3\cr 
                     kb_1(1+\Delta_\tau) & kb_2(1+\Delta_\tau)
                      & kb_3(1+\Delta_\tau)\cr
                       b_1 & b_2 & b_3

\end{pmatrix}
\label{rg2}
\end{equation}
\noindent
where we consider ${(1-\Delta_\tau)}^{-1}\approx 1+\Delta_\tau$ since $\Delta_\tau$ is far less than unity. 
If we consider 
$k(1+\Delta_\tau)\rightarrow k$ then, we get the structure of $m_D$ given 
in eqn.(\ref{mD}). So, $m_D$ with scaling ansatz remains form invariant including RG effect.  
\vskip 0.1in
\noindent
Now, if we consider scaling ansatz breaking through 
$\epsilon$ parameter, the structure of $m_D$ comes out as  
\begin{equation}
m_D = \begin{pmatrix} a_1 & a_2 & a_3\cr 
                     kb_1& kb_2(1+\epsilon)&kb_3\cr
                     b_1 & b_2 & b_3
         \end{pmatrix}.
\label{sbmd}
\end{equation}
Again, RG effect through parameter $\Delta_\tau$ on $m_D$ with broken scaling ansatz is given by 
\begin{equation}
 \begin{pmatrix} 1 & 0 & 0\cr
                      0 & 1 & 0\cr
                      0 & 0 & (1-\Delta)
       \end{pmatrix}
        \begin{pmatrix} a_1 & a_2 & a_3\cr 
                     kb_1& kb_2(1+\epsilon)&kb_3\cr
                     b_1 & b_2 & b_3
         \end{pmatrix}=\begin{pmatrix}   a_1 & a_2 &a_3\cr
                  kb_1& kb_2(1+\epsilon) & kb_3\cr
                  b_1(1-\Delta_\tau) & b_2(1-\Delta_\tau) & b_3(1-\Delta_\tau)

\end{pmatrix}.
\end{equation}
Performing the same exercise of redefinition of $b_i$'s and $k$, the same $m_D$ is obtained as in eq.(\ref{sbmd}). 
So, the structure of $m_D$ matrices with broken scaling ansatz in eq.(\ref{mdsb}) are free from RG effect.
\section{ Expressions used in Sec-4}
In our calculation we have written $m_{\nu}$ by breaking it into two parts, i.e
\begin{equation}
m_{\nu}=m_{\nu}^{0}+\epsilon m_{\nu}^{\prime}.
\end{equation}
If we assume a generic form of $m_{\nu}^{\prime}$ as 
\begin{equation}
m_{\nu}^{\prime}=m_{0} \left( \begin{array}{ccc} A_{1} & B_{1} & C_{1} \\ 
B_{1} & B_{2} & C_{2} \\ C_{1} & C_{2} & C_{3} \end{array}\right)
\end{equation}
(In our case $A_{1}=0$, $B_{1}=kp$, $C_{1}=0$, $B_{2}=2k^{2}p^{2}$, $C_{2}=kp^{2}$, $C_{3}=0$.)\\
The different elements of the matrix $h^{p}$ in terms of the parameters ($k$, $p$, $r$, $\alpha$) are given by
\begin{eqnarray}
&&h^{p}_{11}= m_{0}^{2}(2Re(A_{1})+2kpRe(B_{1})+2pRe(C1))\\
&&h^{p}_{12}= m_{0}^{2}(B_{1}^{\ast}+kpB_{2}^{\ast}+pC_{2}^{\ast}+A_{1}kp+B_{1}k^{2}r e^{-i\alpha} +C_{1}kre^{-i\alpha})\\
&&h^{p}_{13}= m_{0}^{2}(C_{1}^{\ast}+kpC_{2}^{\ast}+pC_{3}^{\ast}+A_{1}p+B_{1}kre^{-i\alpha}+C_{1}re^{-i\alpha})\\
&&h^{p}_{22}= m_{0}^{2}(2kpRe(B_{1})+2k^{2}r(Re(B_{2})\cos \alpha +Im(B_{2})\sin \alpha) +2kr(Re(C_{2})\cos \alpha \nonumber\\
&&+Im(C_{2})\sin \alpha))\\
&&h^{p}_{23}= m_{0}^{2}(kpC_{1}^{\ast}+k^{2}re^{i\alpha}C_{2}^{\ast}+kre^{i\alpha}C_{3}^{\ast}+B_{1}p+B_{2}kre^{-i\alpha}+C_{2}re^{-i\alpha})\\
&&h^{p}_{33}= m_{0}^{2}(2pRe(C_{1})+2kr(Re(C_{2})\cos \alpha +Im(C_{2})\sin \alpha)+2r(Re(C_{3})\cos \alpha\nonumber\\ 
&&+Im(C_{3})\sin \alpha))
\end{eqnarray}
Parameters like $x$, $y$, $z$, etc can be expressed in terms of different elements of $h^{p}$ matrix as
\begin{eqnarray}
&&x=c_{12}(h^{p}_{11}c_{12}-h^{p}_{12}e^{i\phi}c_{23}s_{12}+h^{p}_{13}e^{i\phi}s_{23}s_{12})-c_{23}s_{12}(h^{p^{\ast}}_{12}e^{-i\phi}c_{12}
-c_{23}s_{12}h^{p}_{22}+h^{p}_{23}s_{23}s_{12})\nonumber\\&&+s_{23}s_{12}(h^{p^{\ast}}_{13}e^{-i\phi}c_{12}-h^{p^{\ast}}_{23}e^{-i\phi}c_{23}s_{12}+h^{p}_{33}s_{23}s_{12})\\
&&y=c_{12}(h^{p}_{11}s_{12}+h^{p}_{12}e^{i\phi}c_{12}c_{23}-h^{p}_{13}e^{i\phi}s_{23}c_{12})-c_{23}s_{12}(h^{p^{\ast}}_{12}e^{-i\phi}s_{12}+h^{p}_{22}c_{12}c_{23}
-h^{p}_{23}s_{23}c_{12})\nonumber\\&&+s_{23}s_{12}(h^{p^{\ast}}_{13}e^{-i\phi}s_{12}+h^{p^{\ast}}_{23}e^{-i\phi}c_{12}c_{23}-h^{p}_{33}s_{23}c_{12})\\
&&z=c_{12}(h^{p}_{12}e^{i\phi}s_{23}+h^{p}_{13}e^{i\phi}c_{23})-c_{23}s_{12}(h^{p}_{22}s_{23}+h^{p}_{23}c_{23})
+s_{23}s_{12}(h^{p^{\ast}}_{23}e^{-i\phi}s_{23}+h^{p}_{33}c_{23})\\
&&w=s_{12}(h^{p}_{11}s_{12}+h^{p}_{12}e^{i\phi}c_{12}c_{23}-h^{p}_{13}e^{i\phi}s_{23}c_{12})+c_{12}c_{23}(h^{p^{\ast}}_{12}e^{-i\phi}s_{12}+h^{p}_{22}c_{12}c_{23}-
h^{p}_{23}s_{23}c_{12})\nonumber\\&&-s_{23}c_{12}(h^{p^{\ast}}_{13}e^{-i\phi}s_{12}+h^{p^{\ast}}_{23}e^{-i\phi}c_{12}c_{23}-h^{p}_{33}s_{23}c_{12})\\
&&q=s_{12}(h^{p}_{12}e^{i\phi}s_{23}+h^{p}_{13}e^{i\phi}c_{23})+c_{12}c_{23}(h^{p}_{22}s_{23}+h^{p}_{23}c_{23})-
s_{23}c_{12}(h^{p^{\ast}}_{23}e^{-i\phi}s_{23}+h^{p}_{33}c_{23})\\
\end{eqnarray}
\newpage
\thispagestyle{empty}


\begin{thebibliography}{99}
\bibitem{Gluza} 
  J.~Gluza and R.~Szafron,
  Phys.\ Rev.\ D {\bf 85} (2012) 047701
  [arXiv:1111.7278 [hep-ph]].

\bibitem{He:2011hs} 
  X.~-G.~He and S.~K.~Majee,
  JHEP {\bf 1203}, 023 (2012)
  [arXiv:1111.2293 [hep-ph]].


\bibitem{Mangano} 
  G.~Mangano, G.~Miele, S.~Pastor, O.~Pisanti and S.~Sarikas,
  Phys.\ Lett.\ B {\bf 708} (2012) 1
  [arXiv:1110.4335 [hep-ph]].

\bibitem{Cao:2011cp} 
  Q.~-H.~Cao, S.~Khalil, E.~Ma and H.~Okada,
  Phys.\ Rev.\ D {\bf 84} (2011) 071302 
  [arXiv:1108.0570 [hep-ph]].

\bibitem{Chao:2011sp} 
  W.~Chao and Y.~-j.~Zheng,
  arXiv:1107.0738 [hep-ph].
\bibitem{Meloni:2011fx} 
  D.~Meloni,
  JHEP {\bf 1110} (2011) 010
  [arXiv:1107.0221 [hep-ph]].

\bibitem{Haba:2011nv} 
  N.~Haba and R.~Takahashi,
  Phys.\ Lett.\ B {\bf 702} (2011) 388
  [arXiv:1106.5926 [hep-ph]].

\bibitem{Balantekin:2011ta} 
  A.~B.~Balantekin,
  J.\ Phys.\ Conf.\ Ser.\  {\bf 337} (2012)  012049
  [arXiv:1106.5021 [hep-ph]].

\bibitem{Zhou:2011nu} 
  S.~Zhou,
  Phys.\ Lett.\ B {\bf 704} (2011) 291
  [arXiv:1106.4808 [hep-ph]].

P.~Novella and f.~t.~D.~C.~collaboration,
  arXiv:1105.6079 [hep-ex].
\bibitem{Balantekin:2010sv} 
  A.~B.~Balantekin,
  AIP Conf.\ Proc.\  {\bf 1269} (2010) 195
  [arXiv:1006.2836 [nucl-th]].

\bibitem{GonzalezGarcia:2010er} 
  M.~C.~Gonzalez-Garcia, M.~Maltoni and J.~Salvado,
  JHEP {\bf 1004},(2010) 056
  [arXiv:1001.4524 [hep-ph]].


\bibitem{Jenkins:2008rb} 
  E.~E.~Jenkins and A.~V.~Manohar,
  Phys.\ Lett.\ B {\bf 668} (2008) 210
  [arXiv:0807.4176 [hep-ph]].

\bibitem{Balantekin:2008zm} 
  A.~B.~Balantekin and D.~Yilmaz,
J.\ Phys.\ G G {\bf 35} (2008) 075007
  [arXiv:0804.3345 [hep-ph]].
\bibitem{Barger:2012fx} 
  V.~Barger, R.~Gandhi, P.~Ghoshal, S.~Goswami, D.~Marfatia, S.~Prakash, S.~K.~Raut and S U.~Sankar,
  arXiv:1203.6012 [hep-ph].

\bibitem{Ahn:2012tv}
  Y.~H.~Ahn and S.~K.~Kang,
  arXiv:1203.4185 [hep-ph].
\bibitem{Brahmachari:2012cq} 
  B.~Brahmachari and A.~Raychaudhuri,
  arXiv:1204.5619 [hep-ph].

\bibitem{xx}X.~G.~He and S.~K.~Majee, 
JHEP 1203 (2012) 023 
[arXiv:1111.2293 [hep-ph]].

\bibitem{ema}
H. Ishimori and E. Ma,  arXiv : 1205.0075 [hep-ph]. 
\bibitem{minos1}
[MINOS Collaboration]  L. Whitehead, 
Joint Experimental-Theoretical Seminar (24 June
2011, Fermilab, USA). Websites: theory.fnal.gov/jetp, http://www-numi.fnal.gov/pr plots/ .
\bibitem{minos2}
[MINOS Collaboration]  P. Adamson et al., 
[arXiv:1108.0015 [hep-ex]].
\bibitem{t2k} [T2K Collaboration] K.~Abe et al,
  Phys.\ Rev.\ Lett.\  {\bf 107} (2011) 041801
  [arXiv:1106.2822 [hep-ex]].

\bibitem{dc}
{H. De. Kerrect, Low Nu 2011, Seoul, South Korea, 
http://workshop.kias.re.kr/lownu11/ }.
\bibitem{DayaBay}
 [DAYA-BAY Collaboration]  F.~P.~An et al.,
  Phys.\ Rev.\ Lett.\  {\bf 108} (2012) 171803
  [arXiv:1203.1669 [hep-ex]].
\bibitem{last}
[RENO Collaboration] J.~K.~Ahn et al., 
arXiv:1204.0626 [hep-ex].
\bibitem{sc1} 
  A.~S.~Joshipura and W.~Rodejohann,
  Phys.\ Lett.\ B {\bf 678} (2009) 276
  [arXiv:0905.2126 [hep-ph]].
\bibitem{sc2}
  R.~N.~Mohapatra and W.~Rodejohann,
  Phys.\ Lett.\ B {\bf 644} (2007) 59
  [hep-ph/0608111].

\bibitem{Blum:2007qm} 
  A.~Blum, R.~N.~Mohapatra and W.~Rodejohann,
  Phys.\ Rev.\ D {\bf 76}, 053003 (2007)
  [arXiv:0706.3801 [hep-ph]].
\bibitem{Obara:2007nb} 
  M.~Obara,
  arXiv:0712.2628 [hep-ph].

\bibitem{Damanik:2007yg} 
  A.~Damanik, M.~Satriawan, Muslim and P.~Anggraita,
  arXiv:0705.3290 [hep-ph].

\bibitem{Goswami:2008rt} 
  S.~Goswami and A.~Watanabe,
  Phys.\ Rev.\ D {\bf 79}, 033004 (2009)
  [arXiv:0807.3438 [hep-ph]].

\bibitem{Grimus:2004cj} 
  W.~Grimus and L.~Lavoura,
  J.\ Phys.\ G G {\bf 31}, 683 (2005)
  [hep-ph/0410279].

\bibitem{Berger:2006zw} 
  M.~S.~Berger and S.~Santana,
  Phys.\ Rev.\ D {\bf 74}, 113007 (2006)
  [hep-ph/0609176].

\bibitem{sc3}
  S.~Goswami, S.~Khan and W.~Rodejohann,
  Phys.\ Lett.\ B {\bf 680} (2009) 255
  [arXiv:0905.2739 [hep-ph]].
\bibitem{4zero1}
  G.~C.~Branco, D.~Emmanuel-Costa, M.~N.~Rebelo and P.~Roy,
  Phys.\ Rev.\ D {\bf 77} (2008) 053011
  [arXiv:0712.0774 [hep-ph]].
\bibitem{Choubey:2008tb} 
  S.~Choubey, W.~Rodejohann and P.~Roy,
  Nucl.\ Phys.\ B {\bf 808}, 272 (2009)
  [Erratum-ibid.\  {\bf 818}, 136 (2009)]
  [arXiv:0807.4289 [hep-ph]].

\bibitem{4zero2}
  B.~Adhikary, A.~Ghosal and P.~Roy,
  JHEP {\bf 0910} (2009) 040
  [arXiv:0908.2686 [hep-ph]].
\bibitem{4zero3}
  B.~Adhikary, A.~Ghosal and P.~Roy,
  JCAP {\bf 1101} (2011) 025
  [arXiv:1009.2635 [hep-ph]].

\bibitem{4zero4}
  B.~Adhikary, A.~Ghosal and P.~Roy,
  Mod.\ Phys.\ Lett.\ A {\bf 26} (2011) 2427
  [arXiv:1103.0665 [hep-ph]].
\bibitem{rslts1}
  H.~Fritzsch, Z.~-z.~Xing and S.~Zhou,
  JHEP {\bf 1109} (2011) 083
  [arXiv:1108.4534 [hep-ph]].

\bibitem{rslts2}
 M.~Maltoni and T.~Schwetz,
  PoS IDM {\bf 2008} (2008) 072
  [arXiv:0812.3161 [hep-ph]].

\bibitem{rslts3}
 A.~Damanik,
  arXiv:1201.2747 [hep-ph].

\bibitem{sm1}
  S.~A.~Thomas, F.~B.~Abdalla and O.~Lahav,
  Phys.\ Rev.\ Lett.\  {\bf 105} (2010) 031301
  [arXiv:0911.5291 [astro-ph.CO]].
\bibitem{sm2}
  M.~C.~Gonzalez-Garcia, M.~Maltoni and J.~Salvado,
  JHEP {\bf 1008} (2010) 117
  [arXiv:1006.3795 [hep-ph]].
\bibitem{sm3}
S. Parke, {\it Unreavelling the neutrino mysteries: Present and future}, Summary talk at NUFACT10, 
http://www.info.tifr.res.in/nufact2010/procedings.php/. 
\bibitem{mnubb}
 J.~J.~Gomez-Cadenas, J.~Martin-Albo, M.~Sorel, P.~Ferrario, F.~Monrabal, J.~Munoz-Vidal, P.~Novella and A.~Poves,
  JCAP {\bf 1106} (2011) 007
  [arXiv:1010.5112 [hep-ex]].
\bibitem{Covi:1996wh} 
  L.~Covi, E.~Roulet and F.~Vissani,
  Phys.\ Lett.\ B {\bf 384} (1996) 169
  [hep-ph/9605319].
\bibitem{Nielsen:2001fy} 
  H.~B.~Nielsen and Y.~Takanishi,
  Phys.\ Lett.\ B {\bf 507} (2001) 241
  [hep-ph/0101307].

\bibitem{Pilaftsis:2003gt} 
  A.~Pilaftsis and T.~E.~J.~Underwood,
  Nucl.\ Phys.\ B {\bf 692} (2004) 303
  [hep-ph/0309342].
\bibitem{Antusch:2006cw} 
  S.~Antusch, S.~F.~King and A.~Riotto,
  JCAP {\bf 0611} (2006) 011
  [hep-ph/0609038].

\bibitem{Roos:1994fz} 
  M.~Roos,{\it Introduction to cosmology},
  Chichester, UK: Wiley (2003) p 279.

\bibitem{Harvey:1990qw}
  J.~A.~Harvey and M.~S.~Turner,
  Phys.\ Rev.\  D  {\bf 42} (1990) 3344.

\bibitem{Kolb} E.~W.~Kolb and M.~S.~Turner, {\it The Early Universe} 
(Addison-Wesley, Redwood City, CA, U.S.A,) (1990).
\bibitem{Giudice:2003jh} 
  G.~F.~Giudice, A.~Notari, M.~Raidal, A.~Riotto and A.~Strumia,
  Nucl.\, Phys.\ B, {\bf 685} (2004) 89
  [hep-ph/0310123].
\bibitem{Buchmuller:2004nz} 
  W.~Buchmuller, P.~Di Bari and M.~Plumacher,
  Annals Phys. {\bf 315} (2005) 305
  [hep-ph/0401240].
\bibitem{Abada:2006ea} 
  A.~Abada, S.~Davidson, A.~Ibarra, F.~-X.~Josse-Michaux, M.~Losada and A.~Riotto,
  JHEP {\bf 0609} (2006) 010
  [hep-ph/0605281].
\bibitem{bary1}
[WMAP Collaboration], D. N. Spergel et al.,
Astrophys. J. Suppl. 148 (2003) 175 [arXiv:0302209][SPIRES].
\bibitem{bary2}
[SDSS Collaboration] M. Tegmark et al., 
Phys. Rev. D 69, (2004) 103501 [arXiv:0310723][SPIRES].
\bibitem{bary3}
[WMAP Collaboration] C. L. Bennett et al., 
Astrophys. J. Suppl. 148 (2003) 1 [arXiv:0302207][SPIRES].
\bibitem{rg1} 
A.~Dighe, S.~Goswami and P.~Roy,
  Phys.\ Rev.\ D {\bf 73}, 071301 (2006)
  [hep-ph/0602062].
\bibitem{rg2} 
  A.~Dighe, S.~Goswami and P.~Roy,
  Phys.\ Rev.\ D {\bf 76}, 096005 (2007)
  [arXiv:0704.3735 [hep-ph]].
\end{thebibliography}
\end{document}